# Effect of Confinement on Solvent-driven Infiltration of Polymer (SIP) into Nanoparticle Packings

Neha Manohar, Kathleen J. Stebe,* and Daeyeon Lee*

Department of Chemical and Biomolecular Engineering, University of Pennsylvania, Philadelphia, Pennsylvania, 19104, United States

**ABSTRACT:** Nanocomposite films containing a high volume fraction (> 50vol%) of nanoparticles (NPs) in a polymer matrix are promising for their functionality and use as structural coatings, and also provide a unique platform to understand polymer behavior under strong confinement. Previously, we developed a novel technique to fabricate such nanocomposites at room temperature using solvent-driven infiltration of polymer (SIP) into NP packings. In the SIP process, a bilayer made of an underlying polymer film and a dense packing of NPs is exposed to solvent vapor which induces condensation of the solvent into the voids of the packing. The condensed solvent plasticizes the underlying polymer film, inducing polymer infiltration into the solvent-filled voids in the NP packing. In this work, we study the effect of confinement on the kinetics of SIP and the final partitioning of polymer into the interstices of the NP packing. We find that, while the dynamics of infiltration during SIP are strongly dependent on confinement, the final extent of infiltration is independent of confinement. The time for infiltration obeys a power law with confinement, as defined by the ratio of the chain size and the pore size. Qualitatively, the observed time scale is attributed to changes in concentration regimes as infiltration proceeds, which lead to shifting characteristic length scales in the system over time. When the concentration in the pore exceeds the critical overlap concentration, the characteristic length scale of the polymer is no longer that of the entire chain, but rather the correlation length, which is smaller than the pore size. Therefore, at long times, the extent of infiltration is independent of the confinement ratio. Furthermore, favorable surface interactions between the polymer and the nanoparticles enhance partitioning into the NP packing.

## INTRODUCTION

Nanocomposite films composed of inorganic nanoparticles in a polymer matrix provide important routes to form functional materials, as they combine the functionality of the nanoparticles and toughness and processability of the polymer. Composite films with high loadings (> 50vol%) of nanoparticles (NPs) are of particular interest due to their superior mechanical and transport properties; however, such nanocomposites are challenging to make using conventional blending approaches due to the thermodynamic incompatibility of the two materials and high elasticity of such mixtures. Surface functionalization or polymer grafting onto NPs are intensively investigated strategies to improve the dispersibility of NPs in a polymer matrix, but, even with these modifications, volume fractions of NPs greater than 50 vol% are rarely attained.[1–4]

A method to bypass this hurdle is to form polymer-infiltrated nanoparticle films (PINFs), which relies on preassembled nanostructures that are subsequently filled with polymer. Such composites are used as novel functional materials in energy storage,[5-9] separations,[10,11] and optical devices.[12-14] Their superior mechanical properties also enable their use as protective structural coatings.[15-20] Currently available methods for PINF fabrication include melt processing,[16] infiltration from bulk solution,[6,10,12,15,18] initiated chemical vapor deposition (iCVD),[9] and in situ polymerization.[14,17]

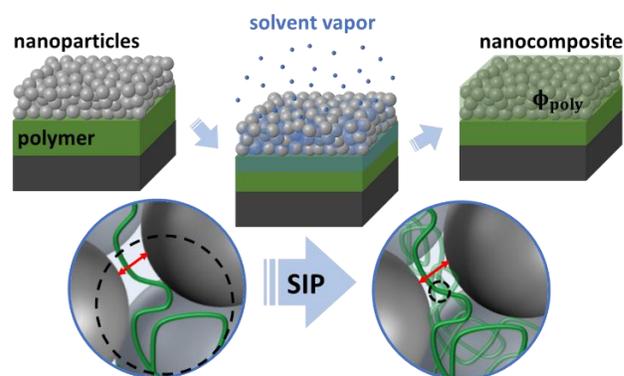

**Figure 1.** Schematic of solvent-driven infiltration of polymer (SIP) into NP packings. The vapor solvent condenses within the NP packing, leading to swelling and infiltration of the underlying polymer film. As the concentration of polymer within the nanoparticle packing increases, the relevant length scales defining confinement within the system change as well. The red arrows indicate the constant pore size, while the black dashed circles indicate shifting polymer characteristic lengths, from the $R_g$ (left) at early times to the correlation length, $\xi$ (right) at later times.

To form PINFs via a scalable route, our group has recently developed the process of capillary rise infiltration (CaRI), in which a polymer infiltrates into the voids of a pre-assembled layer of densely packed NPs via capillarity.[13,19] This process is achieved by heating a bilayer thin film of polymer beneath a layer of densely packed NPs above the glass transition temperature ($T_g$) of the polymer. Capillarity-driven wicking of the polymer melt into the interstitial voids of the NP packing ensues. The CaRI technique requires systems where the $T_g$ is accessible[21,22] and settings in which an increase in temperature does not affect the quality of the NP assembly.

As a complementary method to induce polymer infiltration into the interstices of NP packing, we have recently presented solvent-driven infiltration of polymer (SIP) into nanoparticle packings. In this method, a bilayer film composed of an underlying polymer layer and a densely packed nanoparticle layer is exposed to solvent vapor, leading to capillary condensation of solvent in the interstitial voids of the nanoparticle packing. This condensed solvent subsequently plasticizes the polymer layer below, and leads to infiltration of polymer into the nanoparticle layer (Figure 1).[23] The SIP technique is exciting as a potentially scalable and versatile method to fabricate PINFs with very high nanoparticle content, and can also be utilized to prepare nanoblends of polymer in nanoparticle packings.[24] Moreover, this room temperature method can be particularly useful to prepare PINFs with very high $T_g$ polymers. To enable widespread utilization of this method, it is critical to understand the dynamics and underlying mechanism of this technique.

Here, we explore the dynamics of polymer infiltration and partitioning of polymer into the interstices of NP packing to gain insight on how confinement affects SIP. Different molecular weights and particle sizes are used to probe a range of confinement ratios under a good solvent condition (polystyrene (PS) in toluene, Flory-Huggins parameter, $\chi$ = 0.34). This SIP system provides a unique platform to investigate the effect of extreme nanoconfinement on the behavior of highly solvated polymers. Although the partitioning of polymer from bulk solution to a pore and the relative mobility of polymer within a confining pore have been studied, the concentration regimes have been limited to dilute solutions and solutions just above the critical overlap concentration.[25-28] However, prior studies are lacking in regimes where the characteristic size of the polymer chains is significantly larger than that of confinement, and the concentration of confined polymer transitions from dilute to concentrated regimes.

We show that the kinetics of polymer infiltration into the nanoparticle packing strongly depends on the degree of confinement, whereas the final extent of infiltration into the packing does not. The infiltration time scales strongly with the confinement ratio, which we define as the ratio of the radius of gyration of polymer and the average pore size ($\lambda = R_g/\langle R \rangle_{pore}$). The infiltration dynamics in SIP are slower than chain dynamics predicted for a single chain under cylindrical confinement but faster than the diffusion dynamics in a polymer melt. In contrast, the final volume fraction of polymer within the NP packing is not dependent on the degree of confinement, indicating that at long times, confinement does not hinder infiltration. We postulate that in the SIP system, the environment surrounding the infiltrating polymer shifts over time from dilute to concentrated, leading to changes in relevant polymer length scales, which at long times eliminates the effect of confinement imposed on the polymer chains by the NP packing. We show that the polymer partitions strongly into the voids of the NP packing, which is attributed to the affinity, albeit weak, of PS to the surface of the silicon dioxide ($SiO_2$) NPs.

## EXPERIMENTAL METHODS

***Bilayer fabrication.*** PS of four different molecular weights ($M_n$ = 8,000 g/mol, 80,000 g/mol, 173,000 g/mol and 1,000,000 g/mol; PDI ≤ 1.10) is purchased from Polymer Source, Inc.. $SiO_2$ NPs of two average diameters (LUDOX brand TM-50: $\langle D \rangle_{NP}$ = 22nm, and SM: $\langle D \rangle_{NP}$ = 7 nm) are purchased from Sigma Aldrich. PS solutions are prepared by bath sonication of a PS/ toluene mixture for three hours up to one day (depending on molecular weight), followed by syringe microfiltration (hydrophobic, pore size 0.2 microns). The filtered solution is then spin-coated onto $O_2$ plasma-treated silicon wafers to obtain ~250 nm thickness films. NP dispersions are diluted in DI $H_2O$ and bath sonicated for three hours, followed by syringe microfiltration (pore size 0.45 microns). The NP dispersions are then spin-coated onto the previously prepared PS thin films to obtain a densely packed layer with a thickness of ~250 nm. PS-only thin films with no NP coating are used for in situ swelling studies.

***Scanning electron microscopy (SEM) imaging.*** Cross-section samples are obtained by cleaving pristine or infiltrated samples in half. Top-down and cross-section samples are sputtered with a 4 nm layer of iridium prior to imaging. An FEI Quanta 600 ESEM is used to take scanning electron microscopy images. Images are obtained at 25 kV at high vacuum using an Everhart Thornley detector.

***Ellipsometry measurements.*** A J. A. Woollam alpha-SE spectroscopic ellipsometer is used for in situ and ex situ measurements. In situ measurements are carried out using a vapor annealing chamber setup previously described.[23] Ex situ measurements are carried out by placing samples on a mesh holder suspended directly above a large reservoir of solvent (toluene) within a sealed container. Due to the time it takes for opening and closing this chamber during sample placement, the maximum resolution of the ex situ measurements is ~1 min. The SIP process is quenched by removal of the sample from the chamber followed by placement in a vacuum oven for ~12 hrs. The dried sample is then measured under ambient conditions, using a simple two-layer Cauchy model as previously described.[23]

***Quartz Crystal Microbalance with Dissipation Mode (QCM-D).*** QCM-D measurements are conducted using a QSense E4 system to find shifts in the frequency and dissipation due to mass adsorbing onto an oscillating quartz crystal. The sensor crystal used is a QSX 303 $SiO_2$, which has a 100 nm Au electrode coated with 50 nm of $SiO_2$. Prior to experiments, the crystal is rinsed with distilled water and isopropanol, and then $O_2$ plasma-treated for ten minutes. The crystal is connected to a flow cell with the $SiO_2$ surface exposed, and a baseline frequency and dissipation factor are obtained in air. A wet baseline is collected after flowing toluene over the sensor at 50 sccm for an hour to allow for stabilization. The sensor crystal is then exposed to 0.05wt% polystyrene ($M_n$ = 173,000 g/mol) in toluene solution flowing at 50 sccm. The frequency and dissipation shifts are monitored with continuous flow of the polystyrene solution until stabilization. The sensor crystal is then flushed with pure toluene at 50 sccm for a day to induce desorption of loosely bound polymer. The final shift in frequency and dissipation factor are then modelled using a modified viscoelastic model via the QTools software package (see SI, Fig. S2-S3 for more information). The dry thickness of an irreversibly adsorbed polystyrene layer on a silicon wafer was obtained using ellipsometry measurements. This value is then used to confirm the wet layer density from the QCM-D model, as the dry layer thickness provides an estimate for the volume fraction of polymer in the wet layer.

**Table 1. Characteristics of PS and SiO2 NPs used in this study.**

| $M_n$[a] (kg/mol) | $R_g$[†] (nm) | $\phi^{*}$[b] | $\phi_I^{*}$[‡] | | $\lambda$[c] | |
|---|---|---|---|---|---|---|
| | | | 7 nm | 22 nm | 7 nm | 22 nm |
| 8 | 2.5 | 0.054 | 0.224 | 0.049 | 2.9 | 0.9 |
| 80 | 9.9 | 0.009 | 0.229 | 0.050 | 11 | 3.7 |
| 173 | 16 | 0.005 | 0.230 | 0.050 | 18 | 5.8 |
| 1,000 | 45 | 0.001 | 0.234 | 0.051 | 52 | 16 |

[a]The number average molecular weights ($M_n$) as provided by the supplier; [†]the calculated radius of gyration for unconfined PS in toluene[29]; [b]the unconfined critical overlap volume fraction ($\phi^{*}$); [‡]the calculated confined critical overlap volume fraction ($\phi_I^{*}$) for a polymer in a capillary[30]; [c]the confinement ratio ($\lambda = R_g/\langle R \rangle_{pore}$) for each system considered.[31]

## RESULTS & DISCUSSION

A range of confinement ratios ($\lambda = 1-50$) is studied by varying PS molecular weight and SiO$_2$ NP size. For the systems considered in this study, the radius of gyration ($R_g$), critical overlap concentration in a bulk solution ($\phi^{*}$) and that adjusted for confinement ($\phi_I^{*}$),[30] as well as the confinement ratios ($\lambda$) are provided in Table 1. Only one of the molecular weights ($M_n$ = 8,000 g/mol) is below the entanglement threshold for PS, whereas the other three ($M_n$ = 80,000 g/mol; 173,000 g/mol; and 1,000,000 g/mol) are entangled systems. The radius of gyration of unconfined PS in toluene is determined from the molecular weight using empirical relationships determined via light scattering studies.[29] The NP sizes are chosen such that the curvature is large enough to induce complete capillary condensation within the NP packing; that is, the pores are completely filled with condensed solvent. Two commercially available SiO$_2$ NPs with average diameters of 7 and 22 nm are used in this study. The average pore size within the packing $\langle R \rangle_{pore}$ is estimated from simulations of void volumes in random close packings of spheres.[31] Previous experiments have shown that packings of NPs larger than 70 nm induce partial capillary condensation due to the decrease in the negative curvature of condensed fluid interfaces within the packing.[23,32] SIP occurs in all the systems studied (Fig. 2). The SIP experiments are run to completion, and the time it takes for infiltration to plateau as well as the final value of the plateau are found via ex situ quenching experiments and studied as a function of the confinement ratio, $\lambda$.

In the SIP system, polymer infiltrates from a solvated polymer film, which serves as a concentrated polymer reservoir, into solvent-filled voids between the nanoparticles. Initially, these voids are empty of polymer, and the infiltrating polymer experiences a very dilute environment. As infiltration progresses, however, and more chains enter the voids, the chains begin to overlap and experience more crowding within the voids. Infiltration kinetics therefore reflect polymer passing through various concentration regimes. Figure 3 shows that the time it takes for complete infiltration to occur in each system is strongly dependent on the confinement ratio of the system. This effect cannot be attributed solely to the rate of chain escape from the concentrated reservoir, which would imply an increase in infiltration time with molecular weight as longer chains would require longer times to escape. However, the dependence of infiltration time on pore size indicates that such a time scale cannot be the controlling mechanism, since the same molecular weight of PS takes longer to infiltrate into the packing with smaller pores. For example, in Figure 3, the highest molecular weight PS ($M_n$ = 1,000,000 g/mol) takes roughly one hour to infiltrate into the 22 nm NP packing (dark blue circle, $\lambda$ = 16) but takes ten hours to fully infiltrate the 7 nm NP packing (light blue circle, $\lambda$ = 52). This result indicates that the transport of polymer chains within the solvent-filled pores control the SIP dynamics. As polymer partitions from the crowded swollen film into solvent-rich voids within the packing, the extreme nanoconfinement likely affects the chain conformation and leads to some loss of conformational entropy.

The infiltration time ($\tau$) as a function of confinement ratio can be described by a power law with an exponent of 3/2 ($\tau \sim \lambda^{3/2}$). Although this exponent has not been predicted by an existing theory, prior studies provide scaling for limiting cases. According to a scaling model for infiltration of a single polymer chain into cylindrical confinement, the dynamics are expected to scale with confinement with an exponent of 2/3 ($\tau \sim \lambda^{2/3}$).[34] In the SIP system, however, polymer infiltration eventually progresses into much higher concentration regimes than this model system as the pore is filled with additional polymer chains from the reservoir. The overlap of polymers confined in capillaries is expected to occur at higher concentrations than that of bulk critical ov-erlap due to confined chains occupying less volume than bulk chains (Table 1).[30] When the SIP process nears completion, the concentration of polymer within the pore reaches ∼ 0.7, which is well above the overlap concentration under confinement (Table 1) and very close to the melt state. Thus, the infiltration behavior in the late stage of SIP likely has some resemblance to the dynamics of confined polymer melts. In such systems, infiltration time is predicted to follow a linear relationship with chain length ($\tau \sim N^1$).[33] Since $R_g \sim N^{1/2}$ in an ideal melt, with a confinement ratio defined by the radius of gyration ($\lambda = R_g/\langle R \rangle_{pore}$), the dynamics should scale as $\tau \sim \lambda^2$ (Figure 3). The SIP system passes dynamically from empty to crowded environments within the pore, obeying a power law bounded by the predicted scaling exponents in the dilute and melt regimes. Moreover, interactions with the NP surface could play a role in these dynamics, since the extent of surface coverage presumably evolves as infiltration proceeds. The surface diffusion of polymer melts on attractive surfaces is predicted to have the scaling behavior $\tau \sim N^{5/2}$, while for weak interactions, the relationship is identical to bulk reptation dynamics, namely $\tau \sim N^3$, which both predict a stronger dependence on chain length than that observed in the SIP process.[35]

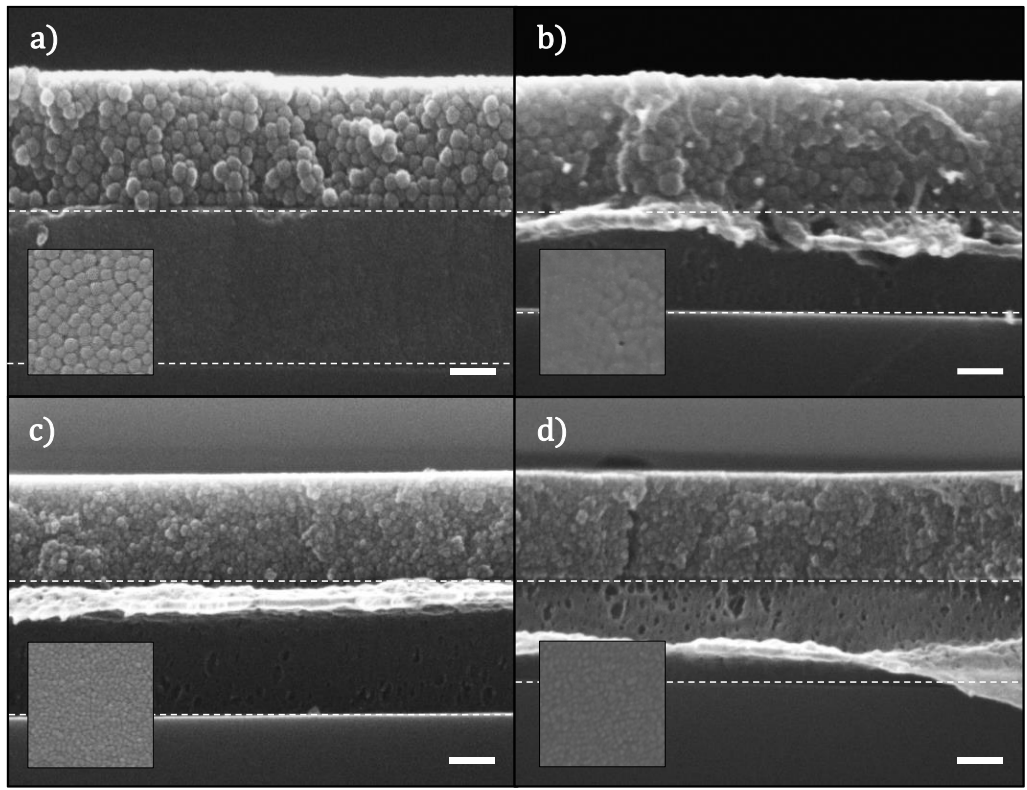

**Figure 2.** SEM images before and after SIP. A) 22 nm SiO2 NPs, PS ($M_n$ = 80,000 g/mol) bilayer before infiltration and B) after infiltration using toluene, C) 7 nm SiO2 NPs, PS PS ($M_n$ = 80,000 g/mol) bilayer before infiltration and D) after infiltration using toluene. Insets show top down images of each nanoparticle packing. Scale bars: 100 nm.

At long times, the polymer eventually stops partitioning into the pore, which is observed experimentally as a plateau in the change in the refractive index of the NP packing and the thickness of the underlying polymer layer. The partitioning of polymer in the SIP systems is defined by two values: the volume fraction of polymer in the external swollen polymer film at equilibrium ($\phi_{p,e}$), and the final volume fraction of polymer inside the voids of the NP packing ($\phi_{p,i}$) (Fig. 4, Fig. S1). To our surprise, all the systems considered in this study infiltrated to the same extent, independent of the confinement ratio, as shown in Figure 4, indicating that confinement has little effect on the final partitioning of polymer into the solvent-filled pore. $\phi_{p,i}$ upon completion of infiltration is approximately 0.7, regardless of $\lambda$. Based on the ratio of this value and the volume fraction of polymer in the swollen film, a partition coefficient $K = \frac{\phi_{p,i}}{\phi_{p,e}}$ for the SIP system can be calculated. The equilibrium volume fraction of polymer in the swollen film ($\phi_{p,e}$), determined from *in situ* ellipsometry of PS films exposed to toluene, is ~0.45 and does not depend strongly on the molecular weight of polymer (Fig. S1). Therefore, the partition coefficient for this system is roughly $K=1.55$, which indicates that despite strong confinement, the polymers prefer to be in the solvent-filled pores than in the solvent-swollen polymer film phase.

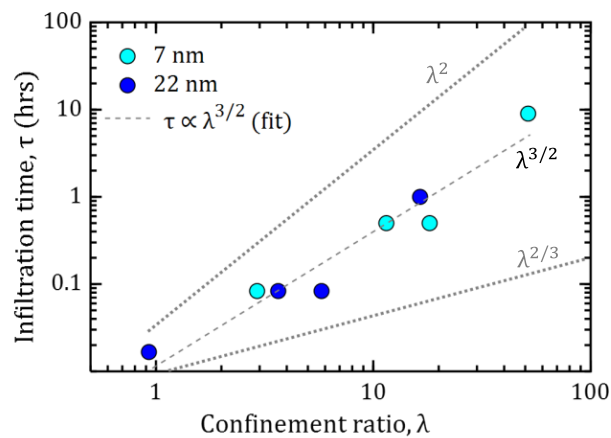

**Figure 3.** Infiltration time (log scale) plotted as a function of confinement ratio (log scale). The dashed line indicates a power law fit, with the critical exponent displayed on the plot ($\lambda^{3/2}$). The infiltration times for bilayers of various MW with SiO2 NPs with $\langle D \rangle_{NP} \approx 7$ nm (light blue circles) and $\langle D \rangle_{NP} \approx 22$ nm (dark blue circles) are shown. The dotted gray lines indicate scaling behavior for melts ($\lambda^2$) and for dilute solutions ($\lambda^{2/3}$).[33,34]

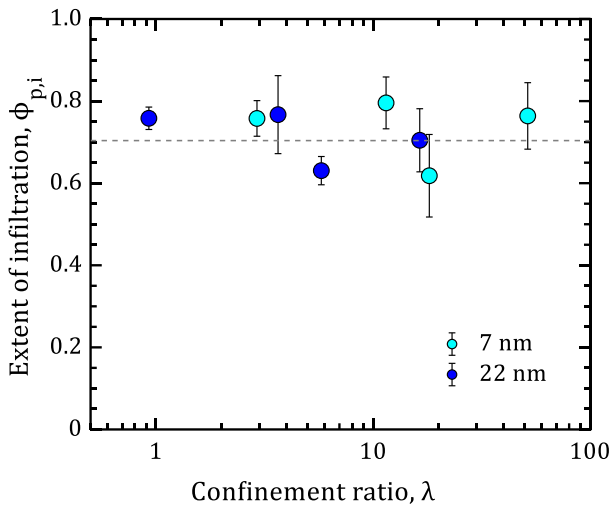

**Figure 4.** The extent of infiltration (volume fraction of polymer in voids of the NP packing) as a function of confinement ratio (log scale). The final extent of infiltration $\phi_{p,i}$ for bilayers of various MW with SiO$_2$ NPs with $\langle D \rangle_{NP} \approx 7$ nm (light blue circles) and $\langle D \rangle_{NP} \approx 22$ nm (dark blue circles) are shown. The value of $\phi_{p,i}$ from SIP does not vary with confinement ratio and has a value ~0.7, indicated by the grey dashed line (partition coefficient, $K \approx 1.55$).

Prior studies have shown that when polymer partitions from a solution phase into a confined space, K is less than 1 and approaches unity as the concentration in the solution phase increases.[25,30,36] This trend is attributed to the osmotic pressure outside the pore at higher concentrations overcoming the entropic barrier of confinement within the pore.[25] However, these studies typically use polymer/pore systems that have purely repulsive interactions. Previous simulation studies indicate that polymer can partition more strongly into confined spaces (K > 1) when there are strong interactions between the polymer and pore surface.[37,38] Although PS has been observed to interact with and adsorb onto unfunctionalized SiO$_2$ surfaces under theta conditions,[39–41] there are a significantly smaller number of studies that investigate the adsorption behavior of PS onto SiO$_2$ surfaces from a very good solvent (toluene). A previous report suggests that some interactions exist between PS and SiO$_2$.[42]

To test whether PS in toluene can adsorb on the surface of SiO$_2$, we perform quartz crystal microbalance with dissipation (QCM-D) experiments. In this experiment, a SiO$_2$-coated quartz crystal is exposed to a continuous flow of dilute PS dissolved in toluene ($M_n$ = 173,000 g/mol, c = 0.05wt%) and the resulting changes in the frequency ($\Delta f$) and dissipation ($\Delta D$) are monitored in situ. We observe substantial changes in f and D when the PS solution is flowing over the crystal, as shown in Figure 5. Once the shifts are stabilized, the sensor crystal is exposed to pure toluene to allow any loosely adsorbed chains to desorb. Even after ~1 day of toluene flow, the frequency and dissipation do not fully recover to the original baseline values. This observation indicates that irreversible adsorption of PS occurs onto the SiO$_2$ surface from the toluene solution. Modelling of this final adsorbed layer using an extended viscoelastic model indicates that the solvated adsorbed layer is roughly 12 nm thick, which is comparable to the $R_g$ of the PS in solution (Fig. S2). It has been postulated that interactions between PS and the untreated SiO$_2$ surface could be attributed to SiOH-π hydrogen bonds forming between the phenyl ring of the PS and the silanol groups on the SiO$_2$ surface.[42] Such an interaction has been observed previously in gas adsorption studies of benzene and toluene onto SiO$_2$.[43]

Unlike the dynamics, the partitioning behavior of polymer into the interstitial voids of the nanoparticle packings in the SIP system does not strongly depend on the confinement ratio (Figure 4). We believe that the independence of partitioning on the confinement ratio is attributed to the fact that, at long times, the concentration within the pores exceeds the critical overlap concentration. The polymer chain's characteristic length scale in such a system becomes the length of unperturbed polymer between chain overlaps, known as the correlation length, which is not a function of molecular weight, but rather of the concentration and stiffness (Kuhn length) of polymer. The correlation length of PS at the final concentration ($\xi \approx 2$ nm)[44], which represents the longest unperturbed chain dimension between overlaps with other polymer chains, is likely to be smaller than the effective pore sizes; in essence the polymer does not feel extra confinement even though the $R_g$ is significantly larger than the size of the pore as the concentration of polymer increases above the critical overlap concentration.

## CONCLUSIONS

In summary, we find that the extent of confinement in the SIP system affects the kinetics of polymer infiltration into the nanoparticle packing, but not the final extent of infiltration in a good solvent. The infiltration time is found to scale with confinement ratio ($\tau \sim \lambda^{3/2}$), which is bounded between the behavior of confined polymer melts ($\lambda^2$) and that of confined single polymer chains ($\lambda^{2/3}$). However, the final volume fraction of infiltrated polymer is independent of the extent of confinement, since the fully infiltrated SIP system is at concentrations well above the critical overlap concentra-tion, where the correlation length scale is smaller (ξ ≈ 2 nm) than the average pore sizes within the NP packings and thus dominates the partitioning. Furthermore, the partitioning from the swollen polymer film into the voids of the NP packing is very high (K ≈ 1.55), indicating that some surface adsorption of PS onto the $SiO_2$ NP surface from toluene occurs in the SIP system. This interaction is confirmed qualitatively via QCM-D studies of a PS solution in contact with a $SiO_2$ surface, which shows irreversible adsorption of a thin polymer layer onto the surface.

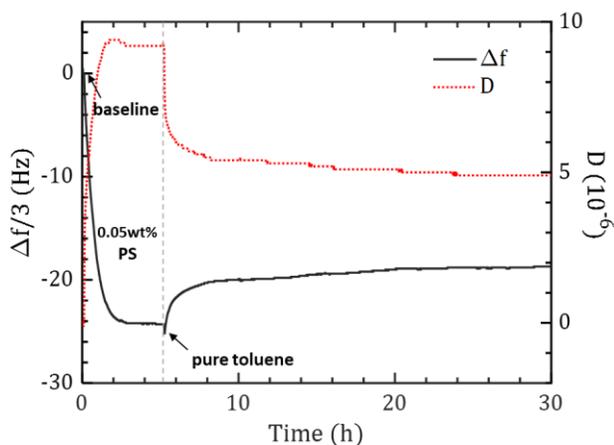

**Figure 5.** Change in frequency (third overtone Δf/3 – black solid line) and dissipative energy loss (D – red dashed line) during PS adsorption and desorption in solution from in situ QCM-D measurements. A 0.05wt% solution of PS ($M_n$ = 173,000 g/mol) dissolved in toluene is exposed to the $SiO_2$-coated sensor crystal (t = 5 min.) and quenched with pure toluene at the point indicated by the grey dashed line (t = 5 h). The frequency and dissipation do not recover to their original values, indicating irreversible adsorption of PS on the $SiO_2$ surface.

Future work will focus on probing how tuning the strength of polymer-surface interaction impacts the dynamics of infiltration in the SIP system and the final extent of infiltration. Molecular dynamics simulations on the SIP system have found that by inducing stronger adhesion between the polymer and NP, a transition from dissolution-dominated infiltration to adhesion-dominated infiltration occurs.[45] Our future work will probe experimentally how altering the nature of the surface interactions affects the extent of infiltration and the infiltration dynamics. To better understand partitioning and kinetics in such systems, an entropic barrier model using self-consistent field theory simulations is currently being investigated.

## ASSOCIATED CONTENT

**Supporting Information**. In situ ellipsometry measurements of PS thin film swelling upon toluene vapor exposure, QCM-D modelling and further PS adsorption experiments using ex situ ellipsometry are provided immediately following the main text.

## AUTHOR INFORMATION

### Corresponding Author

*Email: kstebe@seas.upenn.edu (K.J.S.), daeyeon@seas.upenn.edu (D.L.)

### Author Contributions

The manuscript was written through contributions of all authors. All authors have given approval to the final version of the manuscript.

### Notes

The authors declare no competing financial interest. additional relevant notes should be placed here.

## ACKNOWLEDGMENT

This work was supported by NSF Grant No. CBET-1449337, CBET-1705891 and PIRE-1545884.

SUPPORTING INFORMATION for

# Effect of Confinement on Solvent-driven Infiltration of Polymer (SIP) into Nanoparticle Packings

Neha Manohar, Kathleen J. Stebe*, and Daeyeon Lee*

I.       IN SITU SWELLING EXPERIMENTS OF POLYSTYRENE (PS) THIN FILMS

In order to determine the final swelling ratio of linear atactic PS in the presence of toluene, the swelling behavior of 250 nm PS thin films spin-coated onto silicon wafers is monitored using in situ spectroscopic ellipsometry measurements. In situ measurements are carried out using a vapor annealing chamber setup previously described, at a flow rate of 200 sccm.[1] Since the lowest molecular weight of PS used in the main study, $M_n$ = 8 kg/mol, partially dewets from the substrate when exposed to toluene for greater than 30 minutes, thin films of PS with $M_n$ = 80 kg/mol and 1,000 kg/mol are used to determine if there are any differences due to molecular weight in the final swelling ratio (Figure S1). After exposure for two hours, both molecular weights of PS reach the same equilibrium volume fraction of polymer, $\phi_{p,e} \approx 0.45$.

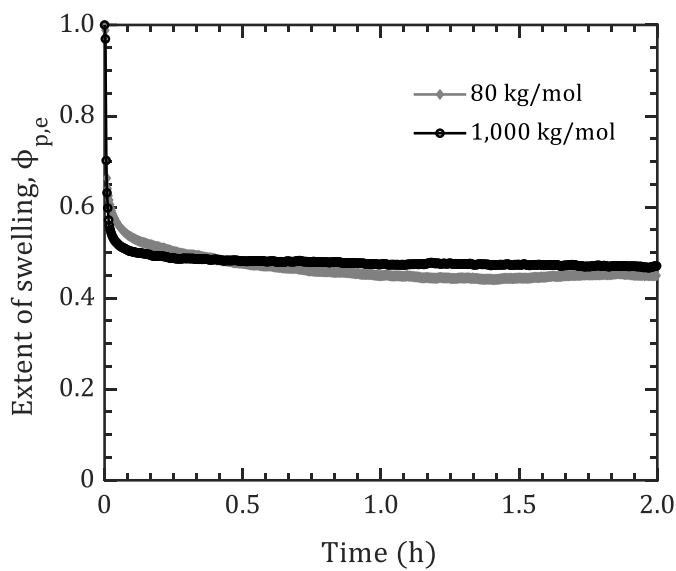

**Figure S1**. Swelling of PS thin films (~250 nm) of two different MWs ($M_n$ = 80,000 g/mol – grey diamonds; $M_n$ = 1,000,000 g/mol – black circles) upon exposure to toluene vapor observed using *in situ* ellipsometry. The equilibrium polymer volume fraction $\phi_{p,e}$ is ~0.45.

## II. MODELLING OF QCM-D ADSORPTION EXPERIMENTS

The final solvated or "wet" layer thickness is found by applying an extended viscoelastic model using the QTools software package (Biolin Scientific, Inc.) to the quartz crystal microbalance with dissipation mode (QCM-D) data presented in the main text (Figure 5) for PS ($M_n$ = 173 kg/mol) adsorption from a 0.05wt% solution in toluene onto a silicon dioxide ($SiO_2$)-coated sensor crystal. A Voigt viscoelastic model with fixed fluid density, fluid viscosity, and adsorbed layer density and with fitted adsorbed layer viscosity, shear, and thickness is used to model the 3rd, 5th, 7th, 9th and 11th harmonic frequency and dissipation data. The model error is reported as a $\chi^2$ value.

Figure S2 shows that the final wet layer thickness after desorption is ~12 nm. In order to obtain this value, a fixed layer density of 900 kg/m³ was used. This density is corroborated using dry layer thickness measurements from ex situ spectroscopic ellipsometry, described in SI Section III.

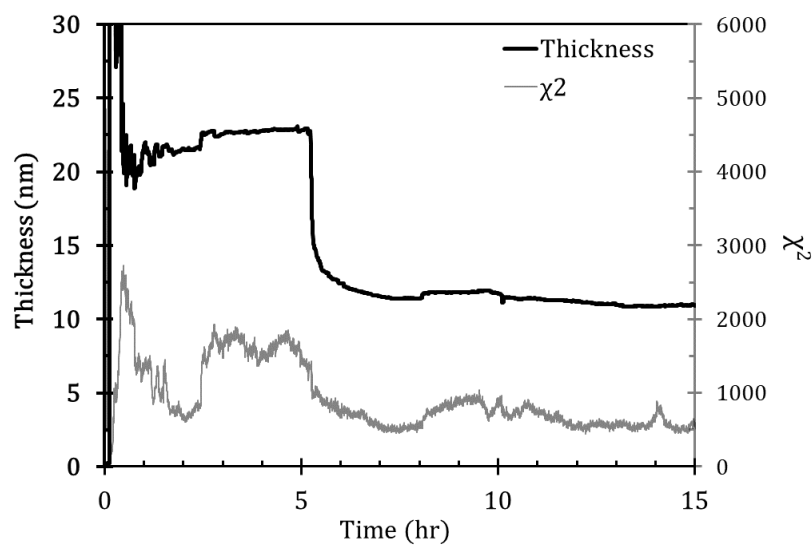

**Figure S2**. The modelled adsorbed layer thickness (thick black line, left y-axis) values for the first ten hours of in situ QCM-D data along with the corresponding $\chi^2$ values (thin grey line, right y-axis).

## III. EX SITU ADSORPTION MEASUREMENTS USING ELLIPSOMETRY

To corroborate the in situ QCM-D measurements, ex situ adsorption studies are performed using spectroscopic ellipsometry. Fresh silicon wafers are cut into 2 cm x 2 cm squares and cleaned with water and isopropanol, followed by oxygen plasma treatment for five minutes to form a native oxide ($SiO_2$) layer. The wafers are then immediately placed inside a 0.05wt% bath of PS ($M_n$ = 173 kg/mol) in toluene. After a given amount of time, the samples are removed from the bath and washed in toluene for five minutes and left in pure toluene for an hour to allow for desorption. The sample is then dried under nitrogen and measured using ellipsometry. A simple Cauchy model is used with oxide ($SiO_2$) layer thickness and angle offset fixed in all measurements based on the values for a bare silicon wafer immediately after oxygen plasma treatment. The polymer layer refractive index is fixed at 1.58, while the thickness is allowed to fit. The results are shown in Figure S3.

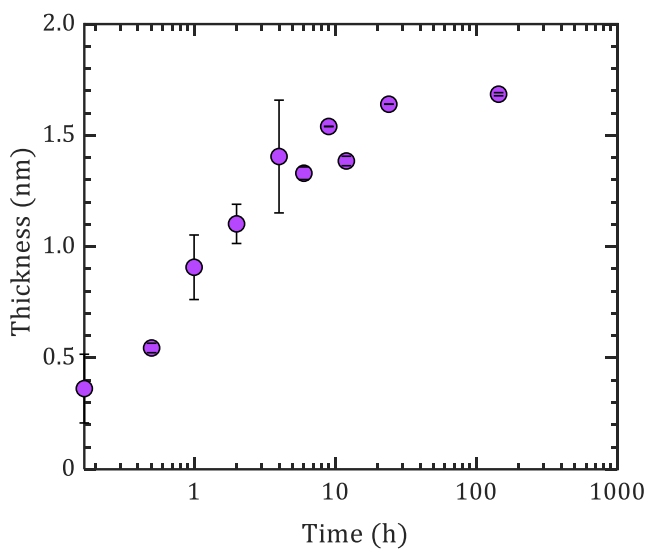

**Figure S3**. The dry adsorbed layer thickness values from ex situ ellipsometry measurements of PS ($M_n$ = 173 kg/mol) adsorbing onto silicon wafer, plotted against exposure time in hours (log scale).

The final thickness of PS adsorbed onto the oxygen plasma-treated silicon wafer is ~1.7 nm. Based on this value, and the value of the wet adsorbed layer of PS from the QCM-D measurements, the wet PS layer density can be estimated. If the volume fraction of polymer in the adsorbed layer is calculated from the dry layer thickness divided by the wet layer thickness, $\phi_{poly} = \frac{\text{Dry thickness}}{\text{Wet thickness}}$, the volume fraction of polymer in the wet layer is ~0.14. With a pure PS density of 1,060 kg/m³ and a pure toluene density of 867 kg/m³, we obtain an estimated solvated or "wet" layer density of 897 kg/m³. This closely matches the fixed adsorbed layer density used in the QCM-D modelling of 900 kg/m³ (see SI Section II).